\documentclass{icrc2009}

\usepackage{graphicx}   
\usepackage{caption}    
\usepackage[font=footnotesize]{subfig} 
\usepackage{fixltx2e}
\usepackage{url}
\usepackage{hyperref}
\newcommand{\shorttitle}[1]%
{\markboth{Proceedings of the 31\MakeLowercase{$^{st}$} ICRC, {\L}\'{o}d\'{z} 2009}{#1} }
\newcommand{\etal}{\MakeLowercase{\textit{et al. }}} 

\begin{document}
\title{Highlights from the Whipple 10-m VHE Blazar Monitoring Program}

\author{\IEEEauthorblockN{Ana Pichel\IEEEauthorrefmark{1}
			  for the VERITAS Collaboration\IEEEauthorrefmark{2}}

                            \\
\IEEEauthorblockA{\IEEEauthorrefmark{1}Instituto de Astronom\'ia y F\'isica del Espacio (IAFE), CONICET, Argentina (anapichel@iafe.uba.ar)}
\IEEEauthorblockA{\IEEEauthorrefmark{2}see R.A. Ong et al (these proceedings) or http://veritas.sao.arizona.edu/conferences/authors?icrc2009\\}
}
\shorttitle{Ana Pichel \etal Whipple Blazar Monitoring Program}
\maketitle

\begin{abstract}
Approximately 25 blazars are known to emit VHE (E$>$100 GeV) gamma rays. Understanding these powerful objects requires long-term, intense, monitoring observations since they exhibit strong, rapid and irregular variability across the entire electromagnetic spectrum. The Whipple 10-m Gamma-ray Telescope, the world$'$s fourth most sensitive VHE telescope, is used primarily to perform such monitoring in the VHE band. The 10-m monitoring program focuses in particular on Mrk 421, Mrk 501, H 1426+428, 1ES 1959+650 and 1ES 2344+514, with observations performed every moonless night that each source is visible. Upon detection of a flare, alerts are sent to VERITAS and the astronomical community to trigger ToO observations, as was the case for one of the brightest-ever VHE flares of Mrk 421 in 2008. In addition to flaring alerts, the 10-m program is used to create long-term light curves, with unprecedented VHE sampling, that can be combined with other multi-frequency observations to better understand blazars. Highlights from the recent Whipple 10-m blazar program will be presented.
  \end{abstract}


\section{Introduction}

Active Galactic Nuclei (AGN) are galaxies whose central cores are believed to be supermassive black holes surrounded by bright accretion disks.
In many cases relativistic jets are produced that accelerate particles to very high energies, producing highly variable emission across the electromagnetic spectrum from radio waves to gamma rays.
Blazars are a subclass of AGN, in which the viewing angle of the jet is very small ($<$$10^{\circ}$), such that the observer is looking straight down the jet and the jet is the most obvious feature of the galaxy.
Blazars are, among AGN, the most powerful astronomical sources known at present. They exhibit strong, rapid and irregular variability over the entire electromagnetic spectrum. Episodes of high variability are produced in a compact zone of the system, most probably in the relativistic jet. The mechanisms responsible for the variability are the motive for this study in very high energy (VHE) gamma ray astronomy.
Blazars provide us the unique opportunity to observe the properties of these processes occurring within the jets. In particular the determination of the types of particles accelerated within the jet is the main question to which we seek answers. To that end detailed spectra across the entire EM spectrum have been measured from blazars in their various high and low states. From these measurements it has been determined that blazars display a wide variety of spectra, and these appear similar to the spectra expected from the Synchrotron Self Compton (SSC) model. Models involving gamma ray emission of hadronic origin can not also be ruled out.

It is essential to have long-term, well sampled, observations of a blazar to understand the emission mechanisms \cite{shell1}. Multiwavelength observations of gamma ray emitting blazars are important in order to test models of non-thermal emission from these objects. Measurements of the temporal correlation between flux variations at different wavelengths during flares are particularly useful, simultaneously providing constraints on the emission models in various energy regimes \cite{shell2}. However, it is difficult to organize a long-term blazar monitoring campaign over different wavelengths due to the long time required to be dedicated only to one source. The VERITAS Collaboration, operating the Whipple 10-m telescope \cite{shell3} and the VERITAS array of new generation gamma ray telescopes \cite{shell4}, is now uniquely positioned to use the Whipple telescope as a long-term monitoring in the very high energy band.

\begin{figure}[!t]

  \centering
  \includegraphics[width=3.0in]{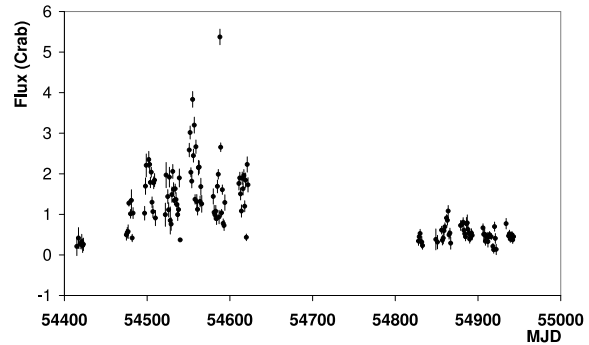}
  \caption{ Light curve for Mrk 421, from September 07 to April 09. The point with the 5.5 Crab flux correspond on May 02, 2008.}
  \label{simp_fig}
\end{figure}

Gamma-ray emission from blazars has been detected in the TeV energies since 1992, when gamma rays from Mrk 421 were observed by the Whipple 10-m Gamma-ray Telescope \cite{shell5}. Whipple detected four other blazars within its sensitivity limits, Mrk 501 \cite{shell6}, H1426+428 \cite{shell7}, 1ES 1959+650 \cite{shell8} and 1ES 2344+514 \cite{shell9}.
\begin{figure*}[!t]
   \centerline{{\includegraphics[width=3.0in]{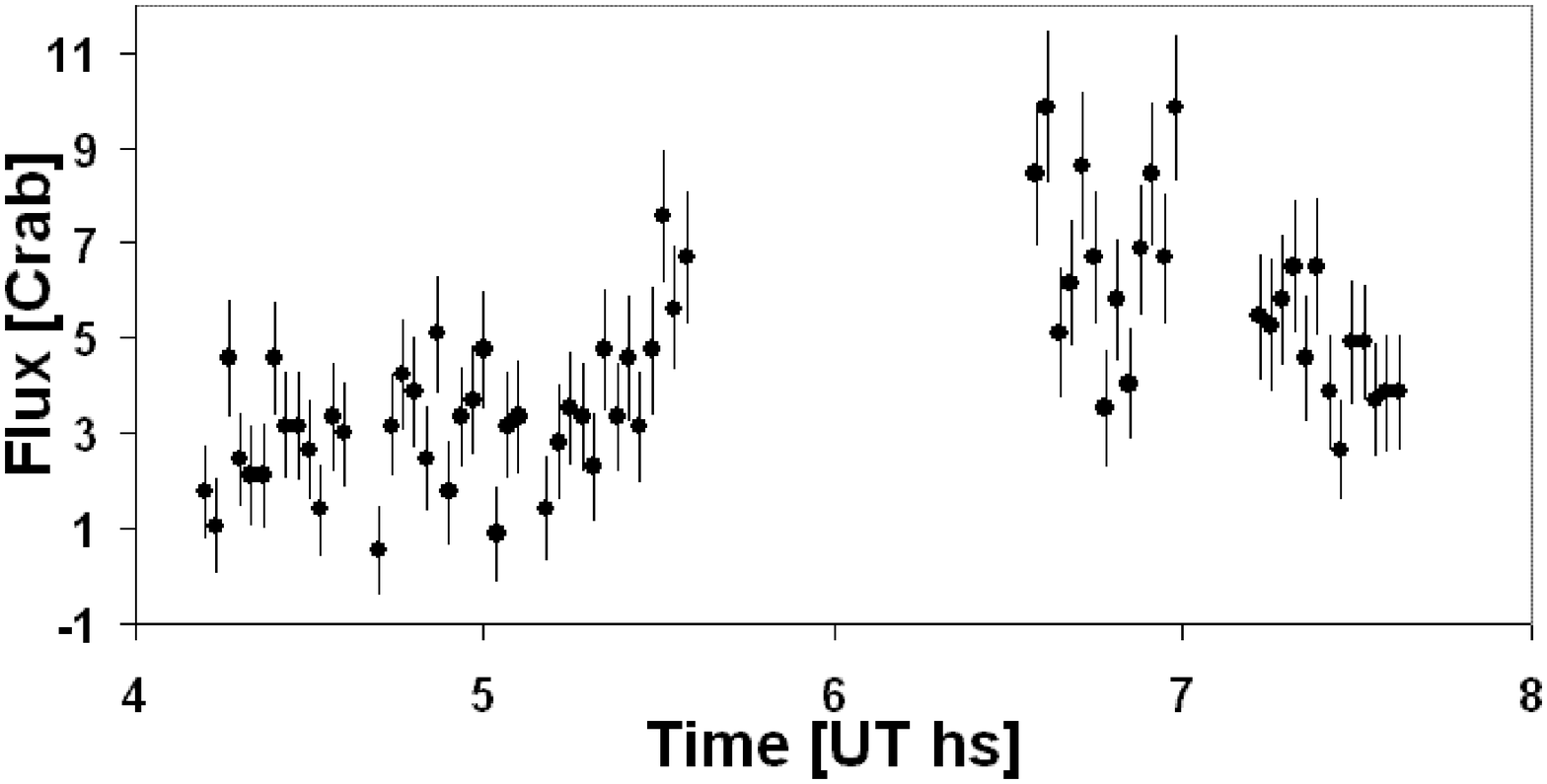} \label{sub_fig1}}
              \hfil
              {\includegraphics[width=3.0in]{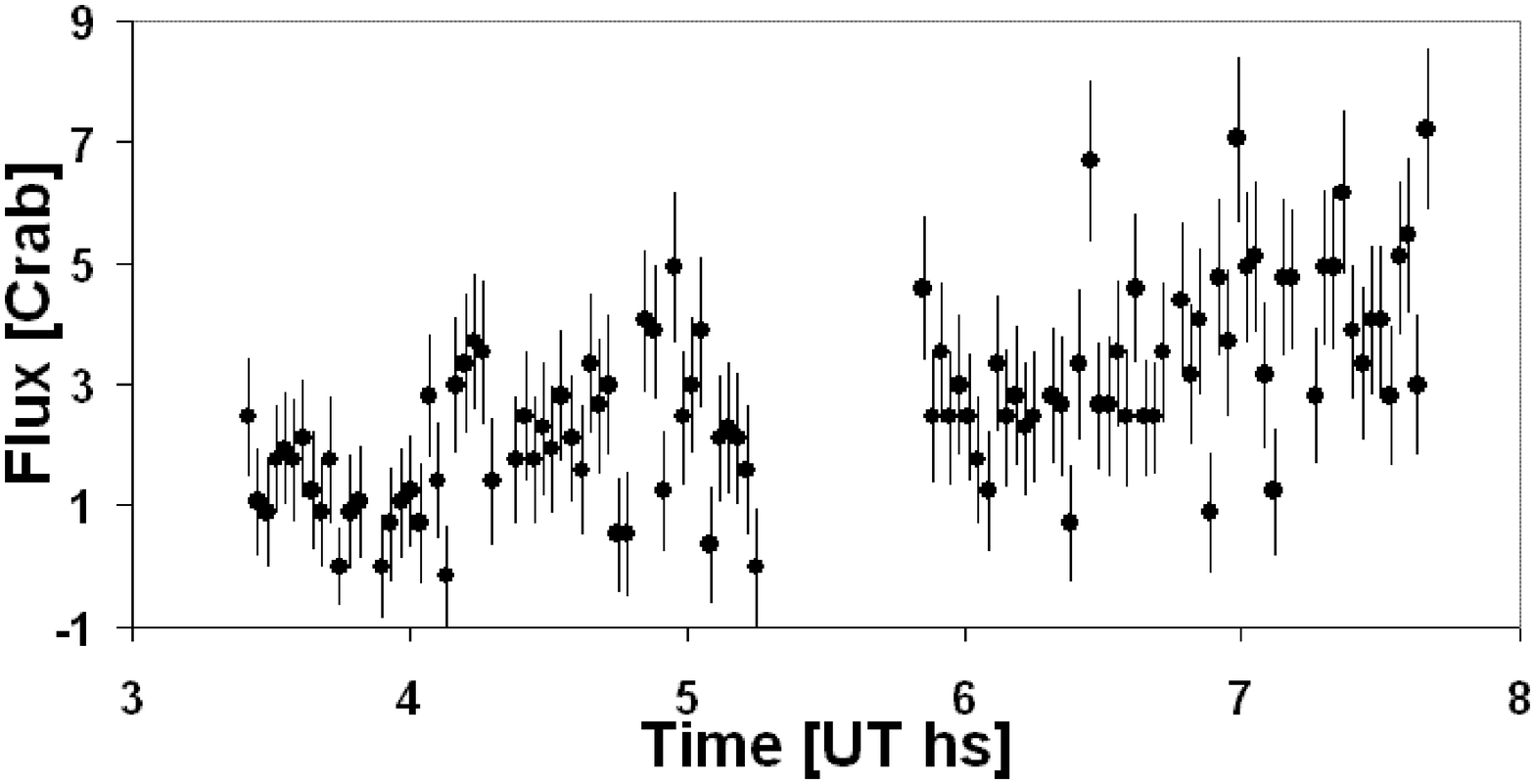} \label{sub_fig2}}
             }
   \caption{Light curves for Mrk 421 taken with the Whipple Telescope. Each point corresponds to a bin time of 2 minutes. {\sl Left}: May 02, 2008. A maximum of intensity around $\sim$10 Crabs is observed.
{\sl Right}: May 03, 2008. A maximum of intensity $\sim$8 Crabs is observed.}
   \label{double_fig}
\end{figure*}
 Since 2005, the Whipple 10 m Gamma-ray Telescope has been used mainly to monitor these five sources each night that they are visible, typically with elevations more than 55 degrees, on cloudless and moonless nights.

Alerts are sent to VERITAS and the astronomical community anytime these objects are flaring to trigger ToO observations. This allows the VERITAS Collaboration to combine with the Whipple long-term monitoring data with the greater energy and time resolution of VERITAS array obtained at this flare phase.
A number of multi wavelength observing campaigns have been undertaken by numerous collaborators with the Whipple program and a significant amount of data has been accumulated \cite{shell2}. We report here on the status of these multi wavelength observations and present light curves of radio, optical, X-ray and gamma ray for some of these objects.

\section{Methods}

All the data taken at TeV energies was registered with the Whipple 10-m Gamma-ray telescope and with the VERITAS telescopes.
The observations are conducted in two modes: ON-OFF and TRK. In the first case, the telescope tracks the source which is centered in the field of view (FOV) for 28 minutes (ON run). The corresponding OFF run is collected at an offset of 30 minutes from the source$'$s right ascension for a period of 28 minutes. The two runs are taken at the same declination over the same range of telescope azimuth and elevation angles. This removes systematic errors that depend on slow changes in the atmosphere. In this mode, direct background substraction of cosmic-ray events in the ON run (as determined by the OFF run) is allowed.
In the TRK mode only ON runs are taken, and with no corresponding OFF observations. TRK mode is useful for monitoring variability of well-established sources. The background is estimated from events whose major axis points away from the center of the FOV. TRK observations are also taken when the sky is possibly cloudy. Clouds in the field of view are noticeable from fluctuations in the cosmic ray rate. An OFF run is, in this case, not possible because the background rate is changing.
Often, data are taken in a mixed mode where ON-OFF runs are interspersed in TRK mode data in such a way that one achieves the best balance between systematic and statistical errors.

\begin{figure}[!t]
  \centering
  \includegraphics[width=3.0in]{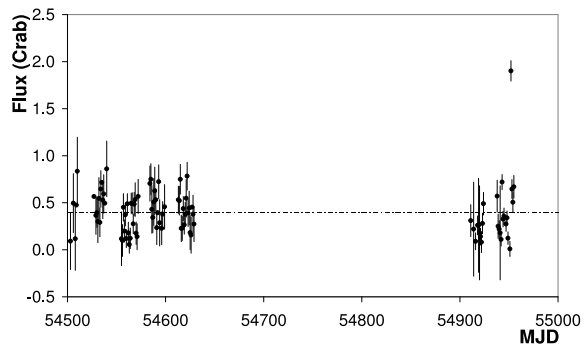}
  \caption{Light curve for Mrk 501 from February 2008 to May 2009. The dotted line is the average flux for the period.}
  \label{lcm5_fig}
\end{figure}

\section{Results and Discussion}

From nightly monitoring of the five sources with the Whipple telescope, we present the results from observations of Mrk 421, Mrk 501 and 1ES2344+514. The final results for the other sources will be presented in future papers.

\subsection{Markarian 421}

Mrk 421 is the brightest blazar detected in X-rays and UV and is the first extragalactic source detected at TeV energies.
The Whipple telescope made observations of 150 hs of Mrk 421 from September 2007 to June 2008 with a significance of 55$\sigma$ in that period of time. For this season (2008-09), the Whipple made a total of 115 hours of observations with a detection of 38.4$\sigma$.
Figure 1 is an example of the long-term monitoring results of the Whipple Blazar program. It shows the data on Mrk 421 over a time period of a year and a half with the Whipple 10-m gamma-ray Telescope.

 A short flare of TeV gamma rays with unusual luminosity from Mrk 421 was observed with the Whipple telescope and triggered VERITAS observations on May 02, 2008.
\begin{figure*}[th]
  \centering
  \includegraphics[width=5.0in]{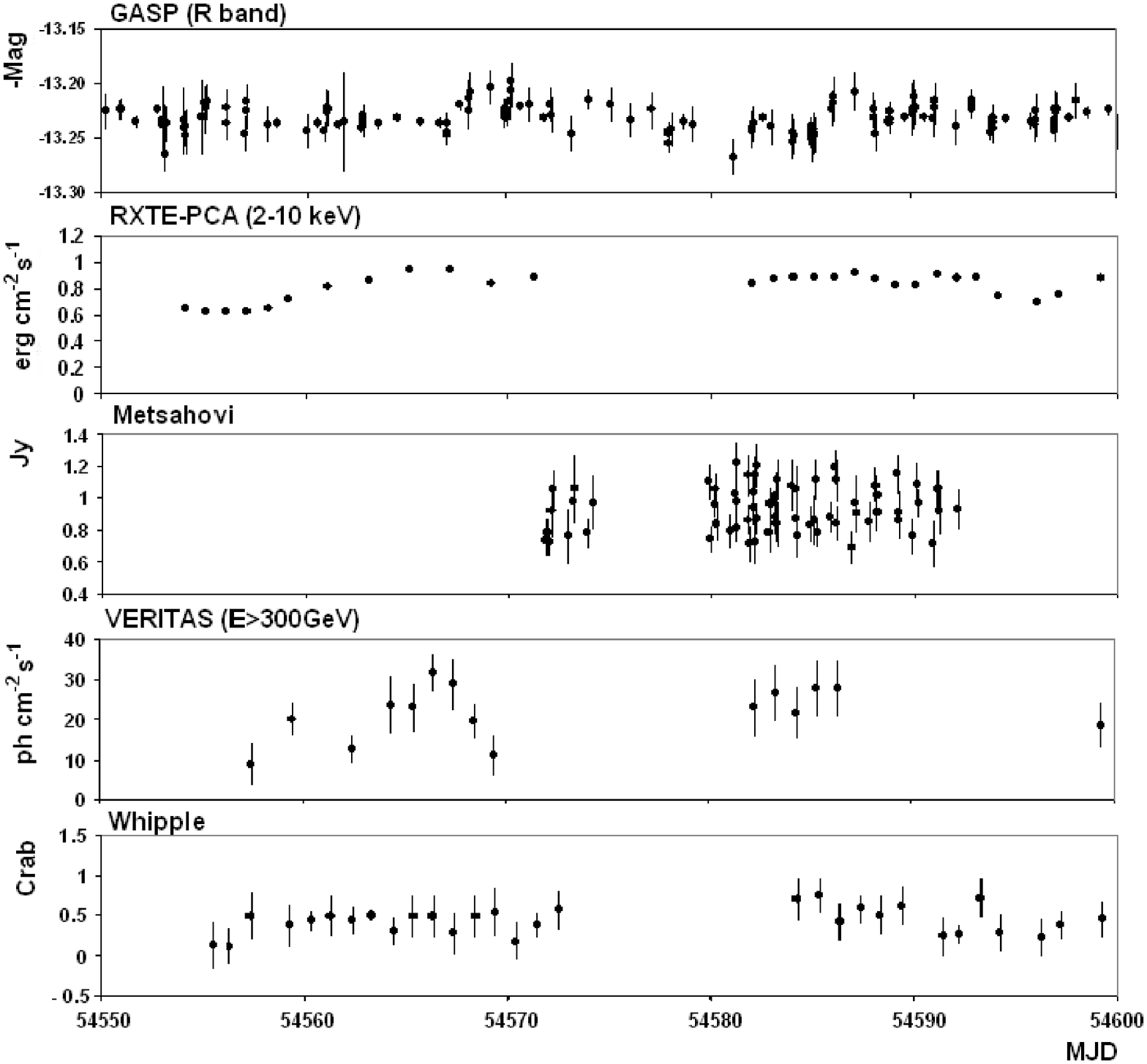}
  \caption{Light curves for Mrk 501 during March-May 2008 corresponding to GASP (optical flux), RXTE (X-ray), Metsahovi (radio band), VERITAS (gamma ray) and Whipple (gamma ray).}
  \label{mwlc_fig}
\end{figure*}
This outburst reached 10 times the flux of the Crab Nebula with a 7$\sigma$ detection. Figure 2 on the left shows the Whipple 10-m light curve for May 2nd and the Figure 2 on the right is the flaring activity on May 3rd for Whipple 10-m.

On May 2 the source showed a steady rise in intensity. After a brief interruption in observations, the source was observed at an intensity of $\sim$10 Crabs, one of the brightest flares observed at the Whipple Observatory. The flux decreased by more than a factor of two in two minutes and then increased by almost the same factor in five minutes. Shortly afterwards the source was observed by VERITAS and the flux was recorded at a level of $\sim$10 Crab \cite{shell10}.

\subsection{Markarian 501}

The Mrk 501 nightly-averaged light curves obtained during the 2008 and 2009 observing seasons are shown in Figure 3.
114 hours of data were taken with the Whipple telescope and resulted in a detection of 11.7$\sigma$, corresponding to 18$\%$ of the Crab rate. The source was at a low state for all this observations except for one night.

On the night of May 01, 2009, Mrk 501 showed a brief period of activity with a high luminosity reaching the 5 Crabs. The Whipple telescope made observations in that night for 2.45 hs with a significance of 19$\sigma$ with a 2 Crab signal over all night. Figure 5 shows the Whipple 10-m light curve for May 01, 2009 where each point corresponds to a bin time of 4 minutes. The spectral energy distribution for the flare will be presented at the conference.

\begin{figure}[!t]
  \centering
  \includegraphics[width=3.0in]{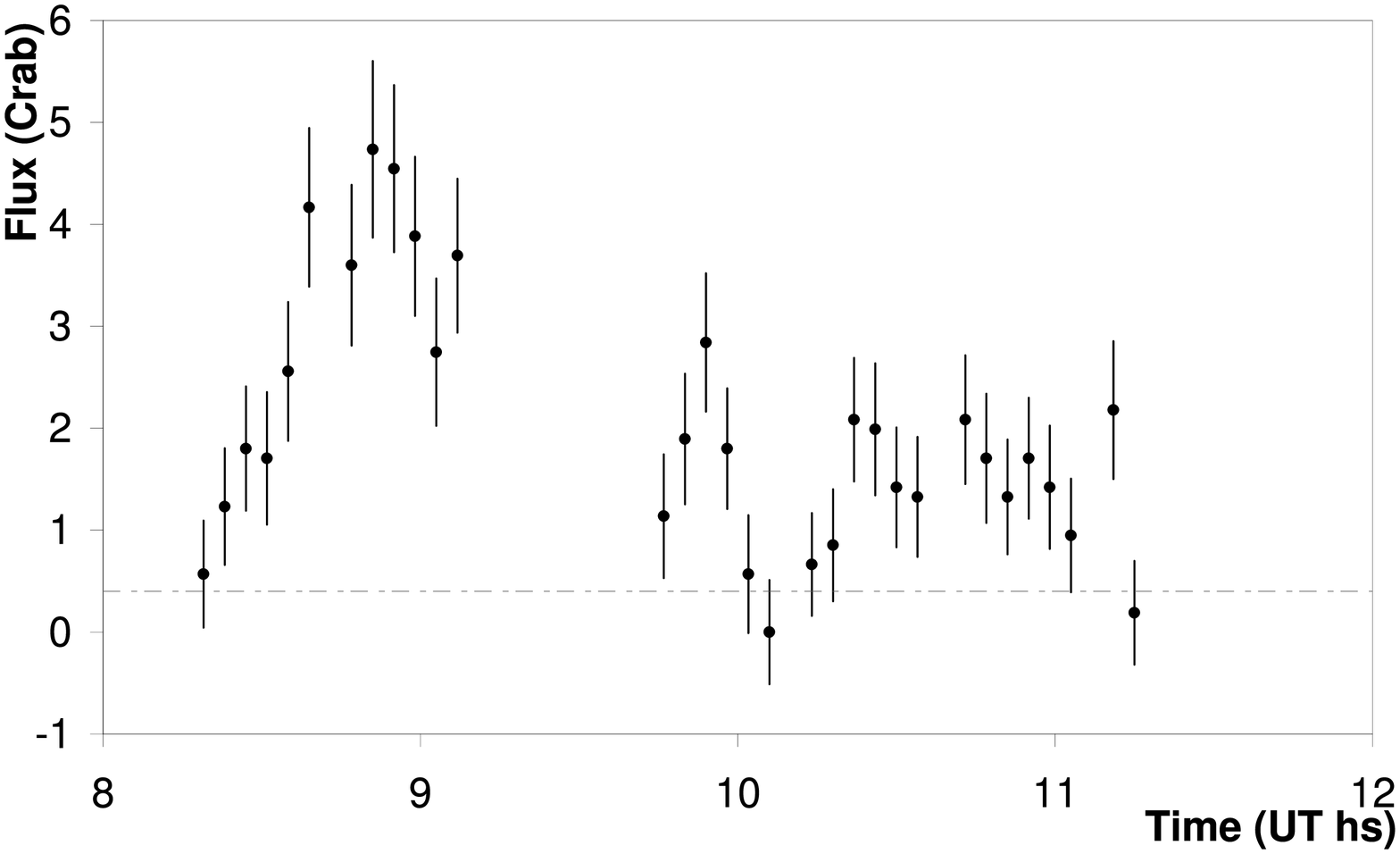}
  \caption{ Light curve for Mrk 501, on May 01, 2009. A maximum of intensity around $\sim$5 Crabs is observed. The dotted line is the average flux for the 2009 season.}
  \label{flaremrk501_fig}
\end{figure}

A multi-wavelength (MWL) campaign from radio to TeV gamma rays was undertaken between March and May 2008. The light curves and spectral energy distribution (SED) analysis of these data is completed, with the data divided into two subsets, low and high states, for the analysis. Figure 4 shows the light curves of GASP in the R band in optical flux, RXTE-PCA in X-ray, Metsahovi in radio band and Whipple and VERITAS in gamma ray.

The source was not very active during this MWL Campaign, being relatively low, although there were very significant flux variations at various frequencies.
Variability seems to increase with energy, the largest flux variations was observed at the highest energies.
The gamma ray and X-ray (2-10 keV) data had, in general, no strong correlation.
A paper detailing the analysis and results will be submitted soon \cite{shell11}.

\subsection{1ES 2344+514}

The nightly VHE gamma ray light curve of 1ES2344+514 from the Whipple telescope is shown in Figure 6. For the nights of November 28 and December 04, 2007 we get a significance of 2.7$\sigma$ for each night. VERITAS detected a strong VHE gamma ray flare on December 7, 2007 corresponding to 48$\%$ of the Crab Nebula flux \cite{shell12}.\linebreak
\linebreak
\linebreak
\linebreak
\linebreak
\linebreak
\linebreak
\linebreak
\linebreak
\linebreak
\linebreak
\linebreak
\linebreak
\linebreak
\linebreak
\linebreak
\linebreak
\linebreak

\begin{figure}[!t]
  \centering
  \includegraphics[width=3.0in]{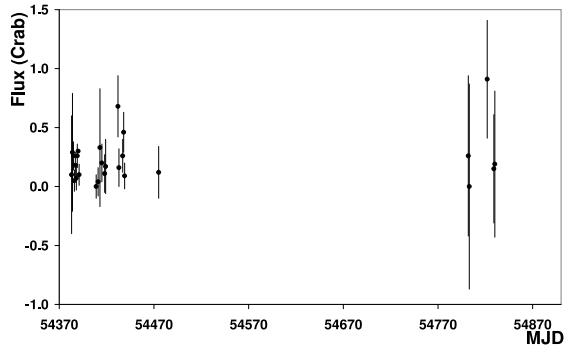}
  \caption{Light curve for 1ES2344+514 with the Whipple telescope from October 2007 to January 2009.}
  \label{lc1es_fig}
\end{figure}

The Whipple telescope was not observing that night due to a bad weather conditionds. Observations with Whipple in VHE gamma rays yield a detection of 4.8$\sigma$ in a total exposure of 58.6 hours live-time \cite{shell13}. 

\section{Conclusions}

These last years of Blazar Monitoring Program with the Whipple 10-m telescope have been very successful in the long-term study of known TeV Blazars. A large number of observatories participated in the program, providing good coverage over a wide range of energies. We have presented multiwavelength observations of the blazar Mrk 501 taken for the last year.
Significant and rapid variability was detected from Mrk 421 and Mrk 501 with the Whipple telescope.
The fourth season of the program is well underway. The sensitive VERITAS \cite{shell4} array is now up and running at the Whipple Observatory and results from 10-m observations are being used to trigger VERITAS AGN observations.

\section{Acknowledgments}

This research is supported by grants from the US Department of Energy, the US National Science Foundation, and the Smithsonian Institution, by NSERC in Canada, by Science Foundation Ireland, and by STFC in the UK. We acknowledge the excellent work of the technical support staff at the FLWO and the collaborating institutions in the construction and operation of the instrument.

\end{document}